\documentclass{caosp306}

\usepackage{graphicx}
\usepackage{natbib}

\articleNo{BP08}
\pubyear{2019}
\volume{49}
\volnumber{2}
\firstpage{367}
\received{November 2, 2018}
\accepted{March 28, 2019}

\begin{document}

\hauthor{ A.\,Kartashova {\it et al.}}

\title{Photometric observations of the asteroid 3200 Phaethon using small and middle telescopes}

\author{ A.\,Kartashova\inst{1}
      \and
        M.\,Hus\'{a}rik\inst{2}
      \and
        O.\,Ivanova\inst{2,3,4}
      \and
        G.\,Kokhirova\inst{5}
      \and
        E.\,Bakanas\inst{1}
      \and
        I.\,Sokolov\inst{6}
      \and
        U.Kh.\,Khamroev\inst{5}
      \and
        A.A.\,Ibragimov\inst{5}
     }

\institute{Institute of Astronomy, Russian Academy of Sciences, Pyatnitskaya str. 48, 119017 Moscow, Russian Federation \email{akartashova@inasan.ru}
         \and
           Astronomical Institute of the Slovak Academy of Sciences, 059\,60 Tatransk\'{a} Lomnica, Slovakia
         \and
           Main Astronomical Observatory of the NAS of Ukraine, 27 Akademika Zabolotnoho str., 03143 Kyiv, Ukraine
           \and
           Astronomical Observatory, Taras Shevchenko National University of Kyiv, Ukraine
           \and
          Institute of Astrophysics of the Academy of Sciences of the Republic of Tajikistan, St. Bukhoro 22, 734042, Dushanbe, Tajikistan
           \and
          Terskol Branch of Institute of Astronomy, Russian Academy of Sciences, Terskol, Elbrus ave., 81-33, Tyrnyauz, Kabardino-Balkaria Republic, 361623 Russian Federation
      }

\date{November 2, 2018}

\maketitle

\begin{abstract}
The main aim of photometrical observations of the asteroid 3200 Phaethon was searching for its low-level cometary activity (possible coma and/or dust tail) in the pre-perihelion passage. We performed observational runs with telescopes ranging from 0.61-m to 2-m and $BVR$ color imaging. Three longer photometric series were used for modeling of the 3D shape of Phaethon. The color indices and size of the asteroid were estimated.
\end{abstract}

\section{Introduction}
\label{intro} Near-Earth asteroid 3200 Phaethon is an outstanding
small body in the Solar System. Its perihelion distance is
0.14\,au, quite close to the Sun, and leads to a systematic
strong heating of the surface. Phaethon reveals cometary activity
\citep{2012AJ....143...66J}, and is considered as a parent body of
the Geminid meteor shower
\citep{1993MNRAS.262..231W, 2019MNRAS.tmp..634R}. These
three features make Phaethon a legitimate target for space-probe
exploration \citep{2018cosp...42E.107A}.

\section{Results from photometry}
\label{phot} Observations of the asteroid 3200 Phaethon were
obtained from October to December 2017 with the 2-m telescope at
the Terskol Observatory (Caucasus, Russia), the 1-m telescope at the
Sanglokh Observatory (Tajikistan), and the 0.61-m telescope at the
Skalnat\'{e} Pleso Observatory (Slovakia). The photometric data
were obtained through the \textit{B}, \textit{V}, and \textit{R}
broadband filters. The reduction of the raw data using bias
subtraction, dark and flat field correction was applied in a
standard way. The color indices, absolute magnitude, and effective
diameter of the asteroid with uncertainties one can see in
Tab.\,\ref{apk_t1}. For calculation of the effective diameter we applied the
formula
\begin{equation}
\log p_V = 6.259 - 2 \log D - 0.4 H,
\end{equation}
where $D$ is the diameter of the asteroid in kilometers, $p_V$ the
geometric albedo, and $H$ the absolute $V$-band magnitude \citep{1989aste.conf..524B}.

\begin{table}[thbp]
    \begin{center}
        \caption{Photometric results of 3200 Phaethon from October to December 2017. Table contains heliocentric and geocentric distances and the solar phase angle, respectively. Then color indices, the absolute magnitude, and the effective diameter of the asteroid with uncertainties. For all diameter estimates we used the value of albedo of 0.1066 \citep {2004PDSS...12.....T}. In the last column there are listed the acronyms for Sanglokh,  Skalnat\'{e} Pleso, and Terskol, respectively.}
        \label{apk_t1}
        \small
        \begin{tabular}{ccccrrrrc} \hline\hline
            Date&$r$&$\Delta$&$\alpha$&$B-V$&$V-R$&$H$~~&$D$~~&Obs.\\
            &au &au&deg&mag~~&mag~~&mag~&km~\\
            \hline
            Oct 28&1.640&0.977&34.0&0.69&0.35&14.37&5.52&San\\
                         &&&&$\pm 0.07$&$\pm 0.04$&$\pm 0.03$&$\pm 0.08$&\\
            Oct 29&1.630&0.957&34.0&0.66&0.36&14.35&5.57&San\\
                         &&&&$\pm 0.13$&$\pm 0.06$&$\pm 0.09$&$\pm 0.22$&\\
            Nov 15&1.444&0.603&32.4&$-$~~&0.43&14.42&5.38&SPO\\
                         &&&&$-$~~&$\pm 0.02$&$\pm 0.03$&$\pm 0.08$&\\
            Nov 17&1.422&0.566&31.9&0.59&0.43&14.40&5.43&San\\
                        &&&&$\pm 0.11$&$\pm 0.07$&$\pm 0.06$&$\pm 0.15$&\\
            Nov 22&1.357&0.461&30.4&0.62&0.35&14.36&5.54&SPO\\
                       &&&&$\pm 0.02$&$\pm 0.02$&$\pm 0.02$&$\pm 0.05$&\\
            Nov 23&1.346&0.443&30.1&0.64&0.36&14.37&5.50&SPO\\
                      &&&&$\pm 0.04$&$\pm 0.03$&$\pm 0.04$&$\pm 0.09$&\\
            Nov 27&1.294&0.367&28.3&0.65&0.33&14.35&5.58&SPO\\
                     &&&&$\pm 0.02$&$\pm 0.02$&$\pm 0.02$&$\pm 0.05$&\\
            Dec 02&1.226&0.273&25.0&0.65&0.35&14.33&5.62&Ter\\
                    &&&&$\pm 0.01$&$\pm 0.01$&$\pm 0.01$&$\pm 0.02$&\\
            Dec 13&1.061&0.089&29.6&0.63&0.35&14.22&5.92&Ter\\
                   &&&&$\pm 0.01$&$\pm 0.02$&$\pm 0.02$&$\pm 0.06$&\\
            \hline \hline
        \end{tabular}
\end{center}
\vspace*{-3.mm}
\end{table}

\begin{figure}[thp!]
\centerline{\includegraphics[width=0.95\textwidth]{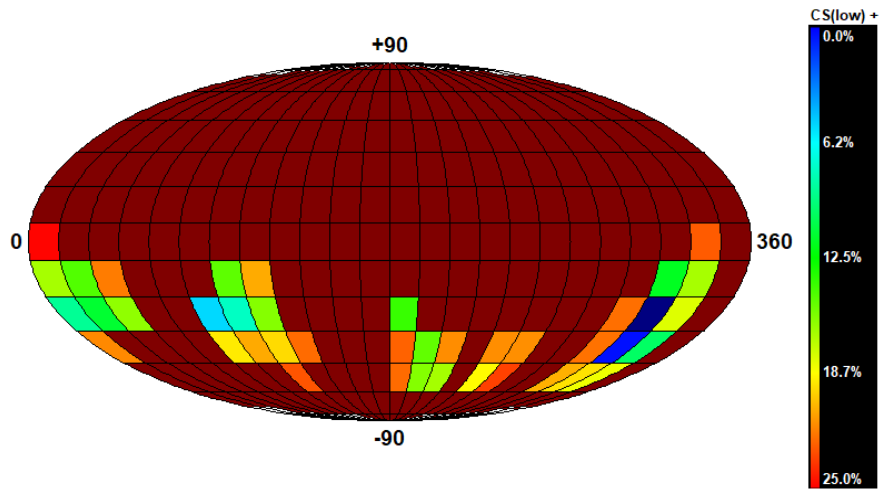}}
\caption{$\chi^2$ residuals between the synthetic and observed lightcurves of the asteroid for spin-vector coordinates covering the entire celestial sphere. The dark blue region represents the pole location with the lowest $\chi^2$, which increases as the color goes from green to yellow and, finally, to red.}
\label{Phaethon-rms}
\end{figure}

\begin{figure}[bhp!]
\centerline{
            \includegraphics[height=3.8cm]{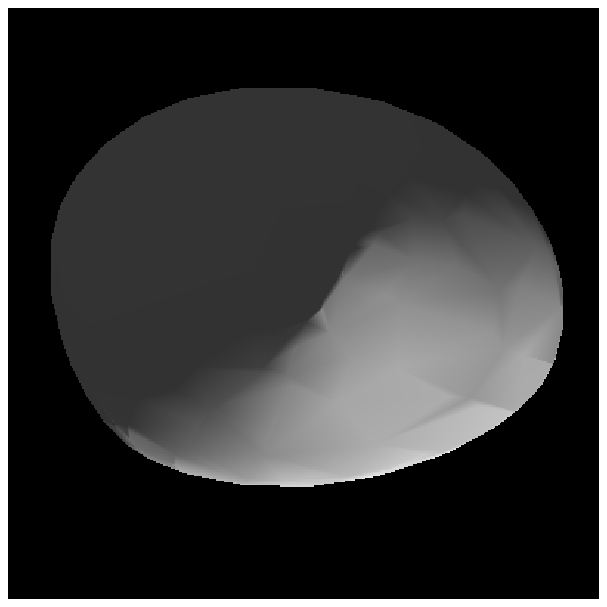}
            \includegraphics[height=3.8cm]{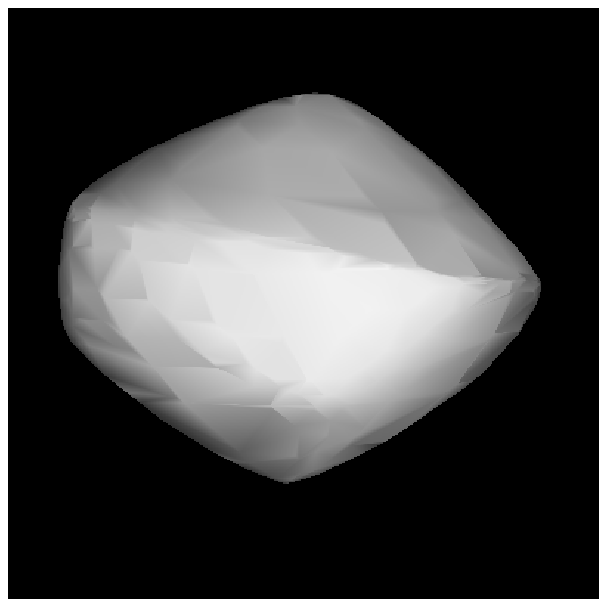}
            \includegraphics[height=3.8cm]{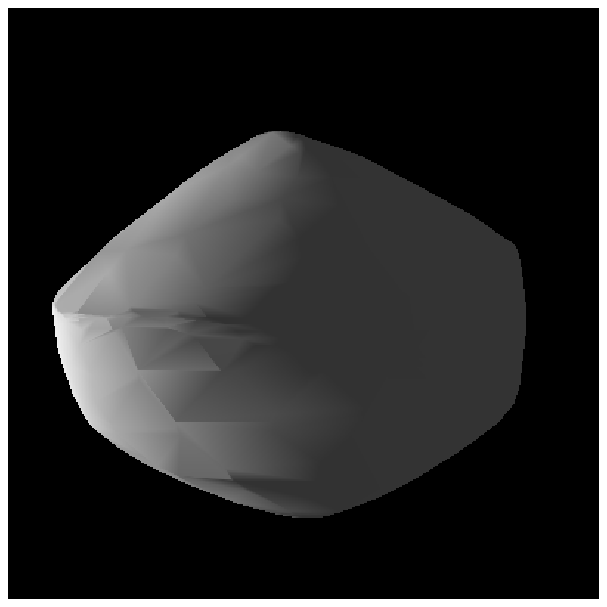}}
\caption{The 3D shape model of Phaethon. On the left there is shown a north pole view, in the middle and right the equatorial viewing and illumination geometry with rotational phases $90^{\circ}$ apart.}
\label{Phaethon_3Dview}
\end{figure}

For comparison our results with other authors we can use the ALCDEF
database\footnote{http://alcdef.org/}. Color indices we can
compare with values from papers of \cite{2012LPICo1667.6294P},
\cite{2013AJ....145..133J}, and \cite{2014ApJ...793...50A}. Our
determined colors are in good agreement with published values.
Absolute magnitudes $H$ have been published by many authors. The values are
in the interval from 13.96 to 14.60\,mag, but only one is reviewed
with three decimal places. It is value 14.345 mag
\citep{2012Icar..221..365P}. Our $H$ values determined in all
nights are very similar to that value.

Also we derived a convex 3D shape model of Phaethon based on 51
previously published in the DAMIT database \citep{2016A&A...592A..34H}
and our 3 lightcurves (November 22 and 27, and December 02, 2017).
We have used the robust method, the so-called lightcurve inversion
described by \cite{KaasalainenTorppa2001} and
\cite{KaasalainenTorppaMuinonen2001}. It allows us to determine
the sidereal period and pole solutions and recover a detailed
convex 3D shape model. The sidereal rotational period was
estimated at the value of $3.603\,96 \pm 0.000\,01$\,hr. This value
is almost the same as computed by \cite{2016A&A...592A..34H} and
reviewed in the ALCDEF database.

Next, the north pole coordinates were computed. In Fig.\,\ref{Phaethon-rms} there is shown the possible location of the north pole at the ecliptic longitude $\lambda_{\rm p}=315^{\circ} \pm 10^{\circ}$ and latitude $\beta_{\rm p}=-31^{\circ} \pm 10^{\circ}$. From that it follows that the sense of rotation of Phaethon is retrograde. The first estimates of the spin axis orientation were published by \cite{2002Icar..158..294K} and \cite{2014ApJ...793...50A}, but the currently accepted, most probable value is that of \cite{2016A&A...592A..34H}. Our estimate is very similar. The latest published position of the north pole of Phaethon at $(318^{\circ} \pm 5^{\circ}, -47^{\circ} \pm 5^{\circ})$ is from the article by \cite{2018A&A...620L...8H}. Our determined 3D shape model has the ratios at $a/b=1.11$ and $b/c=1.04$  (Fig.\,\ref{Phaethon_3Dview}), but that we cannot compare with data from the ALCDEF database.

\section{Conclusion}
Despite many attempts to obtain signs of Phaethon's cometary
activity, no trace of it  has been detected. But our photometry in
all observed nights showed that Phaethon should be actually larger
than 5.1\,km as it was previously published. The analysis from the
Arecibo radar confirmed that Phaethon is really larger and has the
diameter of approx. 6\,km \citep{2018LPI....49.2509T} and an almost
spherical 3D shape\footnote{https://www.jpl.nasa.gov/news/news.php?feature=7030}.
Results of multi-color photometry show a bluish surface for
Phaethon. It is in contrast with typical cometary nuclei that are
slightly reddish in general \citep{2002AJ....123.2070T}.
\cite{2018A&A...620A.179Z} estimated the geometric albedo of
Phaethon to be  $0.075 \pm 0.007$ in the $R$ filter, which appears to
be consistent with dark F-type asteroids.  Also our data confirmed
the taxonomy F-type of Phaethon. Note that Phaethon was classified
as Tholen F-type by \cite{1984PhDT.........3T}.

\acknowledgements AK thanks for the support, in part, from the RFBR
grant No.\,16-02-00805-a. OI is supported, in part, by the project
the SASPRO Program, REA grant agreement No.\,609427, and the
Slovak Academy of Sciences (Grant Vega No.\,2/0023/18). MH thanks
 Grant VEGA No.\,2/0023/18. Observations at the Skal\-nat\'{e}
Pleso Observatory were acquired under the realisation of the
project ITMS No.\,26220120029, based on the supporting operational
Research and development program financed from the European
Regional Development Fund.

\bibliography{ref_apk}

\end{document}